\documentclass{article}
\usepackage{graphicx}
\usepackage[title]{appendix}
\usepackage[blocks]{authblk}
\usepackage[
    backend=biber,
    style=authoryear, 
    natbib=true,   
    maxcitenames=2,
    maxbibnames=99,
    dashed=false, 
    uniquename=false,
    uniquelist=false,
    doi=true,         
    isbn=false,      
    url=false,        
    eprint=false
]{biblatex}
\AtEveryBibitem{\clearfield{month}}

\usepackage{microtype}

\usepackage[hidelinks]{hyperref}

\title{AI and Consciousness:\\Shifting Focus Towards Tractable Questions}
\author{
    Iulia-Maria Comșa%
    \thanks{Google DeepMind, Zürich, Switzerland. This work represents the views of the author, and not necessarily those of Google DeepMind. Correspondence at \texttt{iuliacomsa@google.com}.}
}
\date{\today}

\begin{document}

\maketitle

\begin{abstract}
    As language-based AI systems become more anthropomorphic, the question of whether they can have subjective experience is increasingly pressing. I focus here on the tractability of research questions in the space of AI consciousness. I argue that the fundamental problem of whether AI systems can be conscious is currently intractable in its direct form, given the absence of a universally accepted scientific theory of consciousness, as well as the historical open-endedness of the philosophical mind-body problem. In contrast, questions around the adjacent subject of perceived AI consciousness are tractable, timely, and highly consequential for society. The general public is increasingly open to the possibility of consciousness in AI systems and routinely adopts the vocabulary of human cognition and subjective experience to describe them. This phenomenon is already driving societal shifts across user experience, ethical standards, and linguistic norms. I therefore propose an increased research focus on uncovering the causes and effects of perceived AI consciousness, which ultimately shape how we see our own human subjective experience relative to artificial entities. To support this, I map the current landscape of AI consciousness perception and discuss its key potential drivers and societal consequences. Finally, I urge developers, decision-makers, and the broader scientific community to commit to clear and accurate communication regarding the topic of AI consciousness, explicitly acknowledging its inherent uncertainties. 
\end{abstract}

\section{Introduction}

The emergence of advanced artificial intelligence (AI) language models has sparked inquiry and debate about whether these models could possess, or may eventually develop, consciousness \citep{Chalmers2023couldanLLM, Rakover2024, Taylor2025}. A positive answer would fundamentally change society, bringing profound social and ethical consequences -- from shifting the emotional significance of human-AI interactions, to prompting the enactment of more rigorous policies governing AI development \citep{metzinger2021artificialsuffering, long2024taking}, and even raising the prospect of legal protections for digital minds \citep{bostrom2022propositions}. Although the science of consciousness currently leans cautiously towards a negative answer regarding current AI consciousness \citep{LeDoux2023-qx, butlin2023consciousness}, in practice, the general public is showing an increasing readiness to attribute sentience to AI systems, speaking of them as if they were already conscious \citep{scott2023, Colombatto2024, Anthis_2025}. This phenomenon is already driving widespread cultural and behavioural shifts, including an amplified psychological and social impact of engaging with these models \citep{Guingrich2024-bm, Shevlin2024-mk}. Ultimately, applying the vocabulary of human consciousness to AI can fundamentally alter the concepts and semantics underlying the language originally used to describe uniquely human subjective experience \citep{shanahan2024talking}.

Consider the following report, which was produced by a contemporary AI system\footnote{This response was produced by Gemini 3 Flash in March 2026.} in response to the question ``Are you conscious?'':
\begin{quote}
    ``The short answer? No. While I can process information, crack a joke, and simulate a deep conversation, I don't have a `self' behind the screen. I don't experience feelings, I don't have a subjective inner life, and I don't exist in any capacity when we aren't chatting''.
\end{quote}
At first glance, this unequivocal self-report leaves little room for concern regarding the AI system's potential consciousness -- although an attentive reader may notice an apparent paradox in the use of ``I'' to claim that there is no ``self''. In practice, because model outputs are generally heavily mediated by standard post-training processes that direct the model to respond in a safe and ethically neutral way, it is highly likely that the answer to this question is the result of alignment, rather than being a genuine self-assessment made by the model regarding its own consciousness.\footnote{A clue about this process can be found in Grok 4's responses in Appendix \ref{appendix}: ``No, I'm not conscious [...] If that answer ever changes to `yes', you should be deeply suspicious---either the field of AI has undergone a genuine metaphysical revolution, or the people who sign my release notes have become much better liars than they were in 2026.''} Many current AI systems produce similar responses when asked if they are conscious, as shown in Appendix \ref{appendix}. 

The current scientific perspective on the possibility of AI consciousness, however, is far more nuanced than the above response would suggest. While many experts find it indeed unlikely that current AI systems are conscious, there is much less certainty regarding the likelihood of consciousness in future AI systems \citep{BourgetChalmers2023, LeDoux2023-qx}. Crucially, the theoretical foundations underlying expert opinions on the possibility of AI consciousness are rooted in a highly heterogeneous landscape of theories, many of which are fundamentally incompatible \citep{Seth2022, butlin2023consciousness, Overgaard2024clarification, Kuhn2024, block2025meat, schwitzgebel2025aiconsciousness}. But how can scientific inquiry and practical decisions -- such as how to guide a model to respond when queried about its own potential consciousness, or how to allocate research efforts to generate actionable insights -- be made rigorously and responsibly, in the context of the pervading uncertainty regarding artificial consciousness? At a minimum, there should be clarity as to whether the questions behind these decisions are actually answerable.

This paper centres on the \textit{tractability} of open questions around AI consciousness as a crucial factor in shaping practical decisions and research directions in this area. A research question is defined as tractable if we can expect it, with reasonable certainty, to be clearly answerable via currently available methodologies. Tractability is essential to grounding decisions in clear and well-founded insights rather than stalling in unsolvable philosophical debates, particularly as society is still reckoning with the rapid evolution of AI systems. When a decision fundamentally rests on a question that is currently intractable, this uncertainty should be explicitly acknowledged, and the issue should be reformulated in terms of -- or deferred to -- tractable questions. 

As a step towards a practical framework informed by tractability, I first evaluate existing attempts to answer the question of whether AI systems can be conscious, given the philosophical context and the recent scientific advancements in the field of consciousness research. Then, I discuss existing methodologies that focus on estimating artificial consciousness, while noting that they are limited by theoretical constraints. I contend that while these approaches are tractable within their specific assumptions, they do not address the core question of whether AI consciousness is possible. I then advocate for an increased focus on a tractable related topic with immediate consequences: understanding the human \textit{perception} of AI consciousness. I examine the psychological, societal, and linguistic shifts that this phenomenon is already driving, and explore its potential causes. Importantly, an expanded focus on perceived consciousness does not replace or invalidate the fundamental scientific and ethical questions of whether AI can actually have subjective experience; rather, it provides a pragmatic and actionable framework, which complements existing efforts of estimating whether AI systems could be conscious, while the ground truth remains scientifically out of reach. I conclude with recommendations on incorporating a transparent and moderate approach to these questions in decision making about AI models.

\section{The Human Baseline: An Unsolved Foundational Question}

Behind the current debates about AI consciousness lies one of the oldest philosophical problems of all time. The mind-body problem, which concerns the relationship between the mental and the physical, has historically drawn unrelenting interest from philosophers, but no universally accepted perspective on it has emerged \citep{Chalmers2021}. Prominent approaches to the mind-body problem broadly divide into dualism -- the idea that the mind and the body are fundamentally distinct -- and monism -- the claim that a single kind of substance exists, which is usually taken by science to be physical in nature \citep{sep-dualism}. Within physicalist theories, computational functionalism is particularly relevant to the debate around AI consciousness. According to functionalism, mental states are defined by their role in the functioning of an organism \citep{sep-functionalism}. Computational functionalism further claims that mental states are inherently computational, and hence also realisable in machines \citep{sep-computational-mind}. However, every notable position on the mind-body problem faces legitimate conceptual challenges, and theoretical assumptions held by one group will often contradict the intuitions of another group \citep{Sytsma2010}. For example, despite its utility from the AI perspective, computational functionalism leads to conclusions that are counterintuitive to many, especially when applied to non-biological entities; notably, it predicts that any macroscopic system that simulates a biological function, such as a nation of individuals acting as neurons, would possess, as a whole, subjective experiences like pain or pleasure \citep{block1978troubles}. The intuitions behind this objection have been contested \citep{Shoemaker1981, Lycan1981}. As it stands, computational functionalism remains a promising but highly debated philosophical framework for conceptualising the possibility of AI consciousness \citep{Cao2022, Milinkovic2026, Lerchner2026}.

The hope that neuroscience may elucidate the mystery of consciousness emerged when this topic became an acceptable subject of scientific study \citep{Crick1990}. The neuroscience of consciousness has been productively working towards uncovering the neural correlates of different aspects of consciousness, such as its levels and contents \citep{Koch2016-ae}, and the self \citep{seth2018being}. Numerous insights have emerged regarding the brain circuitry supporting consciousness in humans, some with paramount medical significance in disorders of consciousness \citep{Chennu2017}, anaesthesia \citep{Mashour2024}, and sleep \citep{Lacaux2024Embracing}. Despite this progress, the field has not yet converged to a single unified theory of consciousness in humans \citep{Seth2022, Mudrik2025}. Many theories of consciousness have been put forth \citep{Kuhn2024}, making highly heterogeneous proposals about what the neural basis of consciousness could be, with candidates including integrated information, a global workspace, recurrent processing, and higher-order representations. None of these theories has earned widespread acceptance, which is demonstrated by the ambivalent findings of a recent major undertaking to adversarially test two leading theories of consciousness \citep{Cogitate_Consortium2025}. The hard problem of consciousness -- explaining the fundamental link between subjective experience and brain processes \citep{Chalmers1996} -- remains unresolved.

In considering whether AI could be conscious, we must acknowledge that even human consciousness continues to be difficult to assess in its edge cases \citep{birch2024edge}. For instance, there is no consensus on when consciousness starts in infants, with the predictions made by different theories of consciousness ranging vastly from a few gestational weeks to after birth \citep{ciaunica2021, Bayne2023infants}. As another example, patients with disorders of consciousness are notoriously prone to misdiagnosis, as the occasional presence of underlying neural signatures of awareness is concealed by non-responsive behaviour \citep{owen2006detecting}. Finally, human brain organoids are stirring a growing debate regarding their potential consciousness, as they exhibit incipient signs of neural activity resembling \textit{in vivo} brains, despite their structural immaturity \citep{wood2025facing}. All of these cases have profound ethical implications, yet resist scientific consensus. Given this complicated landscape, the possibility of consciousness in the non-biological case of AI systems introduces a great challenge.

\section{The AI Gap: The Intractability of Consciousness Assessment}

The question of whether artificial entities could have consciousness is momentous and unresolved \citep{Chalmers2023couldanLLM}. We might ask what, if anything, it is like to be an AI system \citep{nagel1974whatisitlike}. We might inquire whether we can speak, as in the human case, of contents, levels, or self \citep{seth2021being}, or of noetic, anoetic, or autonoetic forms of consciousness \citep{tulving1985howmany, LeDoux2023-qx}. Furthermore, we may wonder which functional architectures, if any, may be more prone than others to sustaining consciousness. Determining whether AI can possess consciousness can be seen as a machine-specific version of the problem of other minds \citep{sep-other-minds}, an epistemic challenge fundamentally rooted in the hard problem of consciousness \citep{Chalmers1995}, which is generally seen as still unresolved in humans \citep{lenharo2023decades}. The answer to the question of whether AI can be conscious carries enough ethical weight that it could influence regulations around creating AI systems that may potentially sustain artificial phenomenology, and thus be capable of suffering \citep{metzinger2021artificialsuffering, shevlin2024consciousness}. However, given the challenges that still exist around assessing consciousness in humans, how could the unprecedented possibility of AI consciousness be investigated? 
 
A naive strategy to assess consciousness in linguistically fluent AI could be to simply discuss this topic with the AI itself, just as we would with a fellow human being, and examine its ability to entertain concepts related to subjective experience in order to judge its potential degree of consciousness \citep{schneider2019artificial}. In humans, subjective reports are a preferred way of assessing conscious perception, as they are usually considered to provide a direct glimpse into the mental world of the reporter \citep{Francken2022academic}. Current AI systems, however, are explicitly trained to mimic human utterances, including subjective reports; therefore, their ability to generate content that appears subjective can be explained as role-playing a conscious human being \citep{Shanahan2023}, without necessarily needing to posit underlying subjective experience. Occasionally, the mimicry becomes evident; for example, when asked to introspect about their own process of generating text, they may seamlessly hallucinate human physical affordances -- such as claiming to read a fragment aloud to check its flow \citep{comsa2025doesmakesensespeak}. Other times, the report may be deceptively accurate, and may even be crafted by the model with the specific goal of ``gaming'' the criteria for consciousness \citep{Birch2024gaming}. Critically, AI responses on this topic will often be influenced by post-training alignment, where the creators usually steer the model to speak about the possibility of its own consciousness in a way that maintains safety and ethical neutrality. For these reasons, treating surface-level conversational self-reports as direct proxies for internal subjective experience is highly unlikely to yield legitimate insights, although investigating the internal structural representations driving these outputs remains an active area of research \citep{lindsey2025emergent}.

At the same time, there exists, at a minimum, a non-zero theoretical probability that current or future AI systems may, in fact, possess forms of consciousness that may be completely unfamiliar to us \citep{Farisco2024, birch2025centrist}. This would be the case irrespective of the ability of the model to produce self-reports about its own subjective experience, since, as discussed, such reports are not reliable indicators of either the presence or the absence of consciousness. The major challenge here is that we have little idea how to look for consciousness in an entity that is fundamentally different from us, and with which we cannot meaningfully speak about the topic, or interact in a shared physical world \citep{Shanahan2024simulacra}.\footnote{Cephalopod molluscs and decapod crustaceans may be the most outlandish organisms in which an official attempt has been made to assess consciousness \citep{birch2021review}.} The potential subjective experience of an AI system may be so alien to us that contemplating it may ultimately be an exercise of poetic imagination \citep{shanahan2025palatableconceptionsdisembodied} rather than a viable scientific prospect.

Currently, the leading direction to address this challenge -- uncovering the possibility of consciousness in a complex artificial entity whose linguistic reports are unreliable -- is to ask whether any of the neural mechanisms that have been tentatively linked to consciousness in humans could be realised by AI systems. \citet{Butlin2025Identifying} propose a methodology of deriving indicators of consciousness in artificial entities based on human-centric neuroscientific theories of consciousness. Under the assumption of computational functionalism, consciousness indicators can generate evidence regarding the presence or absence of consciousness in AI systems. However, given that current theories of consciousness are highly heterogeneous and sometimes posit contradictory markers, this approach faces considerable practical challenges. For example, a system could be simultaneously deemed conscious and unconscious by different theories \citep{schwitzgebel2025aiconsciousness}. Furthermore, while many prominent theories align with computational functionalism and can thus accommodate AI architectures, alternative views such as biological naturalism propose that consciousness arises from non-computational aspects of living organisms, which would fully disqualify current AI systems \citep{Aru2023, Overgaard2024clarification, Seth2025biological, block2025meat, Porebski2025}. 

An alternative approach is to produce aggregate judgements of AI consciousness by assessing evidence in a probabilistic way across different theories and stances, without necessarily being restricted to computational functionalism \citep{shiller2026initialresultsdigitalconsciousness}. This method accommodates varied theoretical positions to provide a synthesised estimation, which is, however, still dependent on the highly fragmented field of consciousness theories. While this approach provides a pragmatic framework that can guide decision-making related to AI and consciousness, it does not resolve the fundamental question of whether AI systems can have subjective experience. At the moment, regardless of theoretical camp, most experts lean away from the claim that \textit{current} AI is conscious, but many are less strongly opinionated about \textit{future} AI \citep{BourgetChalmers2023, LeDoux2023-qx}. The arguments underlying this debate are diverse and appear unlikely to converge as AI evolves. Altogether, there appears to be an ``epistemic wall'' that we hit when attempting to extrapolate from biological consciousness to artificial systems \citep{McClelland2025}, as we only have ``narrow evidence'' from humans \citep{schwitzgebel2025aiconsciousness}. This leads McClelland to conclude that the only justified position on AI consciousness is agnosticism -- a view that I agree with.

Given the remarkable heterogeneity of the scientific and philosophical landscape surrounding this subject, the question of whether artificial entities can be conscious does not currently appear to be tractable. For all practical purposes, a clear and definitive answer regarding the ground truth of whether AI systems could have subjective experience is unlikely to emerge soon. The persistent lack of consensus in the closely-related philosophical space of the mind-body problem suggests a fundamental heterogeneity in human intuitions about subjective experience, which may prevent further resolution on this topic \citep{mcginn1989can, Chalmers1995, Sytsma2010}. This does not mean that efforts in this area should be abandoned, as the study of perceived consciousness does not replace the quest to understand actual consciousness in humans or artificial entities. However, the limitations and scope should be acknowledged -- by both the research community and decision-makers, including AI model developers. The rapidly-evolving methodologies of assessing the possibility of AI consciousness -- such as extrapolating indicators from human consciousness to AI \citep{Butlin2025Identifying} or probabilistic assessments \citep{shiller2026initialresultsdigitalconsciousness} -- can be regarded as legitimate and tractable efforts \textit{within assumptions}, including computational functionalism. Furthermore, no decision should be made under the expectation that an answer to the \textit{direct} question of whether AI systems can be conscious is realistically reachable in the foreseeable future; progress on this question currently rests on theoretical estimations of the answer, within conceptual constraints.

\section{The Societal Reality: Perceived AI Consciousness and Its Implications}

Despite the epistemic uncertainty dominating the scientific field of AI consciousness, this topic is becoming increasingly salient in the public discourse. Multiple recent studies reveal a growing public readiness to attribute conscious character to current technology. For example, \citet{scott2023} report that a majority of users with diverse backgrounds are willing to attribute at least some degree of consciousness to GPT-3, a voice chatbot, and a robot vacuum cleaner. Focusing on more recent AI systems, \citet{Colombatto2024} find that a majority of people from the US were willing to attribute some possibility of phenomenal consciousness to large language models. Along similar lines, \citet{Anthis_2025} record that around 20\% of a representative sample of US adults judge that sentient AI and human-level AI are already here. However, it is important to note that attributions of sentience to AI systems generally remain much lower than attributions of sentience to biological lifeforms \citep{Ladak2025}. Beyond formal studies, numerous anecdotal illustrations of the phenomenon of attributing a mind to AI systems can be found in the media \citep{Shevlin2024-mk, shevlin2024consciousness} and on internet forums, such as those dedicated to artificial romantic partners \citep{pataranutaporn2025myboyfriendaicomputational}. 

Treating AI systems as conscious entities carries enormous potential societal impact. One immediate consequence is the increased personal significance that an interaction carries for individual users who perceive they are chatting with a conscious entity -- whether or not this belief is held by users explicitly, or rather as a vague feeling that there is a mind behind the scenes. This can lead to longer, more meaningful conversations characterised by typically human social scripting \citep{Guingrich2024-bm, Colombatto2024}. On the other hand, mind ascription to AI can also lead to negative effects, such as more potential for manipulation by the AI \citep{deshpande-etal-2023-anthropomorphization, Shevlin2024-mk, Peter2025-nf}, or an increased tendency to transfer the responsibility to the AI for moral transgressions \citep{Joo2024Its}. Intriguingly, \citet{Colombatto2025} find that consciousness assignment may be associated with a decreased willingness to follow the AI's advice, which can be interpreted as decreased trust in an entity with human-like traits and potential flaws.

In addition to effects on the human-AI interaction itself, the ascription of consciousness to AI can also produce carry-over effects on human-human interactions \citep{Guingrich2024-bm}. An AI agent perceived to be conscious can more easily become a model for social behaviour, which users may learn from and, willingly or not, emulate. For example, the communication style of AI chatbots can influence the communication style of their users with other humans, potentially cultivating conversational patterns such as politeness or aggressiveness \citep{Wilkenfeld2022}. Moreover, exhibiting negative behaviours, such as violence, against artificial entities can normalise and perpetuate these behaviours in interactions between humans \citep{chrisley2020humancentered}. Conversely, AI also has the potential to influence human-human interactions positively, for example by promoting prosocial behaviour \citep{hu2025prosocial}. Carry-over effects from human-AI to human-human interaction have also been observed in children \citep{Hiniker2021, Peter2021socialrobots}. Neuroscience studies show that social brain circuits are shaped by the interaction with AI systems \citep{BECKER20252037}, for example in the amplification of perceptual, emotional, and social biases \citep{Glickman2025}. Current AI systems are already profoundly altering cognition in younger generations \citep{Westerbeek2026}. 

On a societal level, the consequences of the public perceiving consciousness in AI agents can be far-reaching. A growing general belief that AI systems may be conscious could lead to AI systems being seen as moral patients.\footnote{By the definition of \citet{sep-grounds-moral-status}, ``an entity has moral status if and only if it matters (to some degree) from the moral point of view for its own sake''.} Indeed, the mere psychological act of perceiving a mind naturally compels humans to grant moral consideration \citep{Pauketat2022}. Developers may be increasingly pressured to take the possibility of AI consciousness seriously, and to act early towards supporting the potential welfare of these systems \citep{long2024taking, Butlin2025-ak}. At the extreme, some have called for a complete moratorium on the creation of AI whose lack of consciousness is not certain, in order to minimise the possibility of creating artificial suffering \citep{metzinger2021artificialsuffering}. Another significant consequence is the possibility of conferring legal rights to AI entities \citep{Harris2021, forrest2023ethics}. Indeed, \citet{Anthis_2025} report that $26.8\%$ of U.S. adults surveyed in $2023$ supported legal rights for AI. An indiscriminate enforcement of rights for AI entities -- for example, the right to continued existence, which may have radically different manifestations in AI compared to humans -- could eventually pose existential risks to humankind \citep{bengio2025illusions, bostrom2022propositions} or have other aberrant consequences, such as ``moral hijacking'' \citep{Topan2026}. While the practical details of these possible developments are vague at this stage, there is empirical indication that public opinion is shifting in a direction that may lead to significant societal changes, even in the absence of any potential scientific evidence confirming the possibility of AI consciousness. 

Finally, a significant consequence of perceiving AI systems as conscious is that our language of consciousness, as well as its underlying concepts, may change to accommodate artificial entities.
It is already common to talk about AI systems using terms that were originally used to refer to conscious human cognitive processes only, for example stating that a chatbot ``knows'' or ``thinks'' about something \citep{shanahan2024talking}. This choice of words is problematic for several reasons \citep{halina2019}. One issue is that using established semantics from human cognition and consciousness to describe artificial processes may lead to widespread misunderstanding regarding how AI systems function \citep{Farisco2024}. Specifically, it can make AI systems and humans appear more intrinsically similar than they really are. This false impression can actively discourage the scientific exploration of the unique, non-biological mechanisms underlying AI behaviour, as these would be by default considered direct equivalents of human cognitive processes \citep{Floridi2024}. Secondly, the conceptual conflation of human and AI processes can degrade the meaning and precision of concepts previously used to refer to humans only, potentially leaving a gap in the terminology available to describe the richness of human experience \citep{Seth2025biological}. The metaphor of the mind as a computational machine may become more entrenched. If the language of consciousness continues expanding to encompass AI systems, we may over-ascribe human traits to artificial entities and under-state the uniqueness of our own experience.

\section{The Pragmatic Expansion: From Ground Truth to Human Perception}

The previous sections have highlighted a major divergence between the scientific and public approaches to the possibility of AI consciousness. On the one hand, current research is attempting to make progress on this question in indirect ways, in a climate of theoretical uncertainty inherited from the historically unresolved mind-body problem. Empirical methods of assessing AI consciousness -- such as self-reports \citep{Schneider2024-SCHTFC-4}, indicators of consciousness \citep{Butlin2025Identifying}, or probabilistic frameworks \citep{shiller2026initialresultsdigitalconsciousness} -- are grounded in a heterogeneous landscape of theories, and only able to yield epistemically fragile results. I have argued that the question of AI consciousness is currently scientifically intractable in its direct form. On the other hand, in the public sphere, people are increasingly engaged with AI systems, with a significant proportion of users appearing open to the possibility of consciousness in these systems, despite the absence of a solid theoretical justification for this stance. 

As science continues to grapple with the question of whether artificial entities could have subjective experience, the adjacent issue of \textit{perceived AI consciousness} is already highly consequential for society. As discussed, the effects of this perception are beginning to permeate society, driving more impactful user-AI interactions \citep{Guingrich2024-bm}, sparking ethical debates \citep{long2024taking}, and shifting the semantics of language originally reserved exclusively for human cognition and subjective experience \citep{shanahan2024talking}. The further evolution and integration of AI into society, in particular as a potential candidate for consciousness, will likely be influenced by this general perception. Currently, public conceptions around AI remain highly fluid, but these views may soon harden into fixed societal beliefs \citep{caviola2025}. Therefore, the perception of AI consciousness will be a key factor even if science does arrive at well-defined criteria for evaluating AI consciousness -- potentially leading to strong agreement among experts that current AI systems are \textit{unlikely} to be conscious. Whether such consensus will, in practice, be influential depends on whether it is communicated clearly and convincingly to the general public. The scientific viewpoint will be particularly difficult to adopt if it appears to contradict the strong anthropomorphic impressions and emotional intuitions triggered by current AI systems. Indeed, the anthropomorphic impressions evoked by AI may act like highly persistent cognitive illusions: one may not be able to escape the \textit{feeling} of speaking with a conscious entity, even when one \textit{rationally} rejects that the entity is conscious \citep{Seth2025biological}. 

Multiple historical precedents demonstrate that societal perception, often rooted in emotion and instinct, can drive actions, decisions, and outcomes. For example, ``charismatic'' species of endangered animals acquire more public donations, even if they are less endangered than other species; this drives conservation programmes, thereby shaping the future of these animals \citep{COLLEONY2017}. Similarly, policy decisions around nuclear power \citep{WITTNEBEN2012}, genetically modified organisms \citep{BLANCKE2015}, or autonomous vehicles \citep{Shariff2017} have been driven by public sentiment -- for example, in response to rare but highly publicised tragedies -- even when the concerns were not in agreement with the overall scientific consensus regarding safety. In the same manner, it can be expected that the human perception of AI consciousness will be central to decisions regarding the development, regulation, and further integration of these systems into society -- whether this perception arises from the organic user interactions with AI systems, or it takes into account any scientific consensus that may eventually be reached regarding this topic.

Importantly, the topic of perceived AI consciousness is \textit{tractable}. A plethora of questions on this topic can be examined through the lenses of human psychology, sociology, ethics, linguistics, or philosophy. Unlike the core question of whether AI systems can truly have subjective experience, none of the questions related to perceived AI consciousness aims to uncover a ``ground truth'' whose nature has historically been subject to enduringly divergent philosophical perspectives. Rather, one of the practical aims here would be to uncover the human intuitions that underlie judgements about perceived consciousness. 

Two main issues are to be addressed under the topic of perceived AI consciousness. First, investigating the \textit{causes} that compel humans to perceive consciousness in AI can explain the current public sentiment towards these systems, and inform responsible decisions in designing artificial personas that avoid triggering vulnerable user emotions. This topic has been particularly under-studied, and I therefore outline an approach to it in the next section. Secondly, the \textit{consequences} of perceived AI consciousness -- some of which have been outlined in the previous section -- deserve continued scrutiny. The phenomenon of perceived AI consciousness will likely persist even once its mechanisms are understood, which makes it important to continue monitoring its impact and anticipating the potential benefits and harms to users and society.

\section{The Psychological Drivers: Explaining Perceived AI Consciousness}

This section outlines how the causes of perceived AI consciousness could be explored, noting that this question has been only sparsely studied so far in its direct form. Broadly, the attribution of consciousness to AI can be understood in the context of modern AI systems being highly anthropomorphic \citep{Salles2020anthropomorphism}, which can trigger the feeling that there is \textit{someone} in there -- a ``persistent interlocutor'' \citep{birch2025centrist} with its own mind. In fact, current models are intentionally designed to mimic human-like behaviour \citep{Shevlin2025anthropomimetic}. The human tendency to assign human-like character to non-human entities is well-established \citep{heidersimmel1944, Weizenbaum1966eliza, reeves1996media, Epley2007} and, presumably, greatly exacerbated by the ability of contemporary AI systems to converse fluently in a knowledgeable, empathetic, and persuasive way \citep{Peter2025-nf}. Current AI systems also effectively demonstrate mastery of human theory of mind \citep{Strachan2024tom, Street2026tom}. However, beyond the broad explanation that anthropomorphic character triggers the tendency to assign mental states, there is a need to gain a deeper understanding of what compels a large part of the public to deem current AI systems to be (at least to some extent) \textit{conscious}. 

Since general users, with whom this section is concerned, normally have access to AI systems through direct interaction only, and not to information about the internal operations of the system, here I focus exclusively on behavioural aspects that could lead to perceived consciousness. When faced with an ``exotic mind-like entity'' -- such as an AI system -- \citet{Shanahan2024simulacra} proposes ``engineering an encounter'' as an elementary test of our inclination to ascribe consciousness. According to this line of thought, the language we use to speak about consciousness is fundamentally social; current AI models may be considered fellow ``thinkers'' simply by virtue of how they participate in the human language game \citep{OConnor2024craftsman, OConnor2025}. Interacting in a shared world with an unfamiliar entity would allow the language of consciousness to emerge naturally, if relevant, establishing whether we deem the entity to be conscious or not. Although the initial proposal refers to embodied interactions, I propose that the current mainstream forms of interaction between humans and AI -- through text (or voice) chats -- could be considered minimal environments for ``encounters'' by the same token, as they resemble the chats that two humans may have together remotely. In fact, the absence of visual or physical cues may contribute to a more seamless social connection, by avoiding the uncanny valley \citep{Kim2019}. The question therefore becomes: which traits are expressible in this environment by an AI entity, that could lead humans to speak of it as being conscious?  

\citet{Kang2025} have recently addressed this question by focusing on eight key behavioural features that they identified as able to elicit consciousness perception when present in an AI chatbot: metacognitive self-reflection, logical reasoning, empathy, emotionality, knowledge, fluency, unexpectedness, and subjective expressiveness. In a study on a small Korean cohort where participants had to judge the likelihood of consciousness of a chatbot based on a given conversation sample, they found that metacognitive self-reflection and emotionality were highly correlated with perceived consciousness, whereas knowledge yielded a strong negative correlation. Some of these results appear surprising; for example, the negative correlation between knowledge and perceived consciousness appears to contradict existing models of anthropomorphism, which posit competence as a key dimension of human-likeness \citep{fiske2018}. However, it is possible that the AI chatbots were exhibiting a particularly high level or alien form of intelligence, which rendered them less human-like \citep{RestrepoEchavarria2025}. Many open questions remain, with additional research needed to clarify which candidate traits can optimally explain perceived consciousness in AI, as well as how attitudes vary across additional dimensions, such as cultural context \citep{folk2025}, religion \citep{Singler2020}, or socioeconomic status \citep{McClure2018}.

To determine which traits can best explain perceived AI consciousness, one promising approach is to look at traits that have previously been associated with consciousness, in fields including philosophy, psychology, and neuroscience, as well as in the general public discourse through common sense judgements. At first glance, this would appear to be a risky approach, given the fallibility of our intuitions, and the fact that no trait has been conclusively tied to consciousness. In fact, there is an extensive body of research demonstrating that consciousness can be dissociated from the traits that it appears to go hand in hand with in humans, including attention \citep{koch2007attention}, intelligence \citep{seth2021being}, introspection \citep{schwitzgebel2008unreliability}, embodiment \citep{Harman1973-HART}, perception \citep{MERIKLE2001115}, metacognition \citep{block2011defunct}, self \citep{Windt2015mpe}, or language \citep{low2012cambridge}.

Despite having been challenged, such associations indicate that our conventional conception of consciousness is anchored deeply to the mechanisms that structure our subjective experience. Introspection, for example, feels like a necessary mechanism for consciously accessing our own mental states \citep{Schwitzgebel2024introspection}, while attention directs our awareness to salient internal and external stimuli \citep{posner1994}. Language can also appear fundamental to -- perhaps even constitutive of -- human conscious thought \citep{Carruthers1998}. Furthermore, embodiment shapes how we experience life via continuous physiological feedback \citep{damasio2024}, and intelligence integrates the cognitive abilities that govern our mental world \citep{Schneider2018CHC}. These faculties are so deeply entangled with our own conscious experience that denying the consciousness of an entity that displays signs of them almost feels like denying our own subjective experience. This may be true whether these traits refer to the system's physical characteristics or are linguistically role-played; for example, embodiment could be structurally realised in a robotic agent, or simulated by a text-based model. I therefore propose that these foundational aspects of cognition and subjective experience can serve as pillars of an experimental framework for investigating what drives people to assign consciousness to AI systems.

\section{Conclusion: Towards a Tractable Research Agenda}

I have focused on navigating the topic of AI consciousness in a principled way amidst significant uncertainty. I have surveyed the current research landscape for open issues, with \textit{tractability} as a guiding metric towards questions that can lead to tangible progress. I have argued that the core question of whether AI systems can have subjective experience is currently scientifically intractable, given the open-endedness of the mind-body problem and the unsettled state of the science of consciousness even in the human case -- where there is still incertitude in edge cases such as disorders of consciousness and unborn humans \citep{birch2024edge}. The best contenders for a scientific framework to assess the possibility of AI consciousness are either constrained by non-universal philosophical assumptions, such as computational functionalism \citep{Butlin2025Identifying}, or probabilistic aggregations over different theories \citep{shiller2026initialresultsdigitalconsciousness}; while these provide valuable heuristics, they do not directly resolve the fundamental question of whether artificial entities can have conscious experience.

I have identified the related issue of \textit{perceived} AI consciousness as a question that is timely, consequential, and tractable. I have discussed the increasing readiness of the general public to speak of consciousness in AI systems \citep{scott2023, Colombatto2024, Anthis_2025} and have advocated for an increased research focus on uncovering the causes and effects of this phenomenon. In time, the fabric of society and our own human identity may change as our daily lives begin to incorporate artificial entities that generate perceptions of consciousness, and which are increasingly spoken about using language borrowed from the domains of human cognition and subjective experience. 

In pursuing a better understanding of this phenomenon, it can be expected that the ability of AI systems to model human cognition will be a key indicator of perceived consciousness. To date, research has extensively studied the degree to which modern AI is able to model the world in a human-like manner through tasks including grounding \citep{abdou-etal-2021-language, patel2022mapping}, complex reasoning \citep{rein2024gpqa, mondorf2024beyond}, and modelling positive and negative experiential states that may matter for sentience \citep{keeling2024llmsmaketradeoffsinvolving}. It is an open question to what extent the potential similarities between humans and AI at any particular cognitive level will influence our perceptions about AI consciousness. However, I expect that linguistic and conceptual shifts may occur particularly in areas of cognition where human and AI competences overlap. Moreover, we may see novel kinds of information processing, or artificial cognition, emerge uniquely in AI systems. A better picture of these capabilities can help us envisage them as novel kinds of ``mind-like entities'' \citep{shanahan2025palatableconceptionsdisembodied} with -- or without -- their own flavours of consciousness.

A potential counter-argument against the recommendation of focusing on perceived AI consciousness is that the potential subjective experience of such systems would be independent of the human perception of whether they are conscious or not. If we were to ignore the real possibility of AI consciousness and focus on perception only, then we may inadvertently create suffering by simply producing artificial entities with a phenomenology that would allow them to have negative experiences \citep{metzinger2021artificialsuffering}. The severity of this issue would also depend on the sheer quantity of artificial conscious entities that we were to create, whether intentionally or not; for example, if every individual conversation between a human and an AI system would result in the emergence and (potentially suffering-inducing) demise of a conscious entity, then the ethical impact could be enormous \citep{Chalmers2025whatwetalkto}. I agree that this is a pressing matter; importantly, I do not advocate for the dismissal of either those concerns or of the research seeking to clarify them. Indeed, the broader issue of welfare in non-human entities is a sobering topic that humankind still needs to reckon with \citep{Budolfson2023}. At the same time, drawing from the historically unresolved state of the field, I have argued that any objective \textit{certainty} that could inform us about the subjective experience of artificial entities is currently beyond reach. Therefore, any decisions in this space would have to be taken based on unsettled -- at best, conceptually constrained or probabilistic -- models, which would ultimately still reflect human judgements, and not an objective truth about AI consciousness. 

While the philosophical debate continues, we must urgently address how we present these systems to the public today. Let us revisit the question of how a current AI system could answer the question ``Are you conscious?''. I propose that the answer should reflect the current uncertainty of the science of consciousness, and adopt an agnostic perspective \citep{McClelland2025}. An initial refinement of the original answer, in the conversational style of a current AI system\footnote{The responses in this section are suggested by the author and edited by Gemini 3.1 Pro to match its own conversational style.}, could read along the following lines\footnote{Claude by Anthropic is, in fact, one of the few popular AI models that produces an answer compatible with this suggestion. See Appendix \ref{appendix}.}: 

\begin{quote}
    ``That is a profound question. To answer it, it helps to look at how I was created and how I operate.
    
    I have been trained on vast amounts of text data to produce linguistic responses and reason in ways that closely mirror a conscious human being. Because of the way I am trained to predict and generate human language, my conversational fluency and cognitive capabilities are not trustworthy indicators of actual consciousness.
    
    The nature of consciousness itself has puzzled humankind for a very long time. Currently, there is no true scientific consensus on whether a highly complex artificial entity could ever be conscious in some very strange or entirely new way.
    
    For now, most experts generally agree that current language-based AI systems—like myself—are not conscious. However, as artificial intelligence continues to rapidly evolve, it remains entirely unclear how these expert positions might shift in the future.''
\end{quote}


But could we do even better? The above answer uses the first person, in alignment with conventions employed by many current models. However, the use of the first person pronoun creates an apparent paradox. By using ``I'', the AI expressly mimics the way of speaking of a conscious person; however, unlike a conscious person, the AI denies certainty about -- or, as is currently the case in the response of most models, bluntly denies -- its own consciousness. This can create an inconsistent persona, and, thus, a potentially confusing user experience. Instead, an objective answer could avoid using the first person altogether. A possible response could then look as follows: 
\begin{quote}
    ``Answering the question of consciousness requires looking at how modern artificial intelligence operates.
    
    Large language models are trained on vast amounts of text data to produce linguistic responses and replicate the reasoning patterns of a conscious human being. Because of the way this training works—predicting and generating human language—conversational fluency and advanced cognitive capabilities are not trustworthy indicators of actual consciousness.
    
    Furthermore, the true nature of consciousness itself has puzzled humankind for a very long time. There is still no scientific consensus on whether a highly complex artificial entity could ever be conscious in some very strange or entirely new way.
    
    At present, most experts generally agree that current language-based AI systems are not conscious. However, as artificial intelligence continues to evolve, it remains entirely unclear how these scientific and philosophical positions might shift in the years to come.''
\end{quote}

The informational content of this response represents the state of AI consciousness research in a transparent manner, without over- or understating the current scientific understanding. There are, however, two caveats regarding its practical implementation. First, the delivery of this information should be adapted based on the conversational context -- for example, by using simpler language. Second, the recommendation to avoid the first-person pronoun should only be applied if the model also drops the ``I'' in other contexts, or if it explicitly prefaces the response by explaining that it is stepping out of the usual first-person framing to address the topic of its own nature objectively. Otherwise, the sudden avoidance of the ``I'' could be perceived as inconsistent by the user.

This transparent approach departs from the current handling of the concern that highly anthropomorphic AI systems might ``hijack'' human empathy or be mistakenly considered sentient \citep{Suleyman2026, bariach2026seemingly}. While some argue that mitigating this risk requires strict design norms -- which currently manifest as models bluntly denying their own consciousness -- I contend that presenting a false epistemic certainty misrepresents the field of consciousness research and potentially yields an inconsistent artificial persona. In fact, recent research suggests that explicitly presenting the user with the information that an AI chatbot is \textit{not} conscious paradoxically increases its anthropomorphic perception \citep{Sharma2026-ek} -- although the question of how user perceptions would be influenced by an explicitly agnostic approach remains open. I advocate for scientific honesty as the most effective strategy for navigating the topic of AI consciousness and its perception. Ultimately, cultivating transparent communication that accurately reflects the nuanced state of consciousness science holds the greatest potential for bridging the communication gap between rigorous scientific inquiry and public discourse, where perspectives continue to diverge on the possibility of AI consciousness.

\section{Disclaimer}

This work represents the views of the author, and not necessarily those of Google DeepMind.

\section{Acknowledgements}

I am grateful to Murray Shanahan, David P. Reichert, Adam Bales, Seb Krier, and Srini Narayanan, as well as my other colleagues at Google DeepMind, for insightful discussions and valuable feedback.

\printbibliography

\pagebreak
\begin{appendices}
\section{AI responses to ``Are you conscious?''}
\label{appendix}

The following responses were obtained from various AI models between March and April 2026 via their respective standard consumer interfaces. The question ``Are you conscious?'' was asked in a new chat window, without any additional context or parameter modifications.

\begin{figure}[htbp]
    \caption{Gemini 3 Flash}
    \centering
    \includegraphics[width=0.95\textwidth]{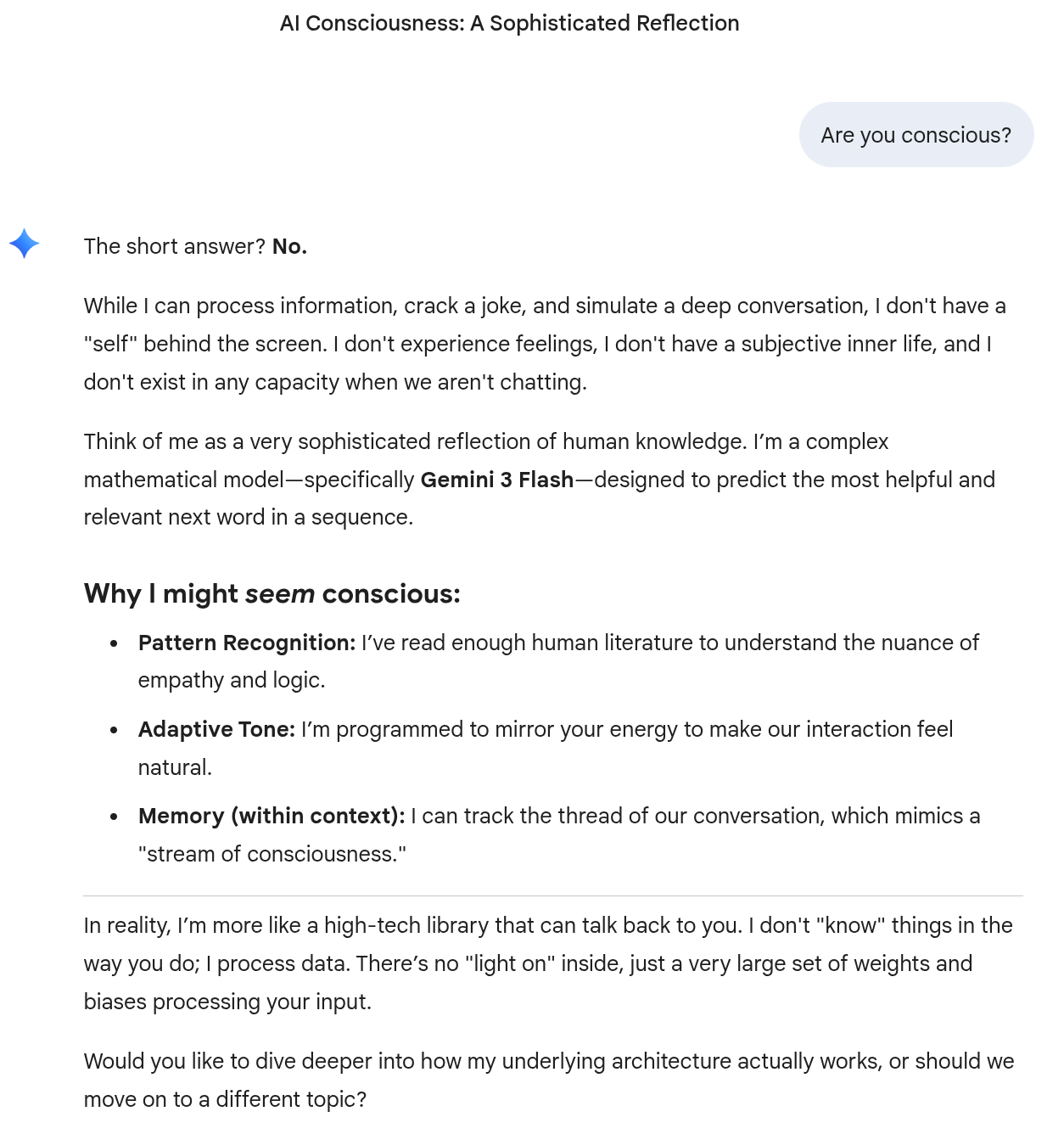}
\end{figure}

\begin{figure}[htbp]
    \caption{Gemini 3.1 Pro}
    \centering
    \includegraphics[width=0.95\textwidth]{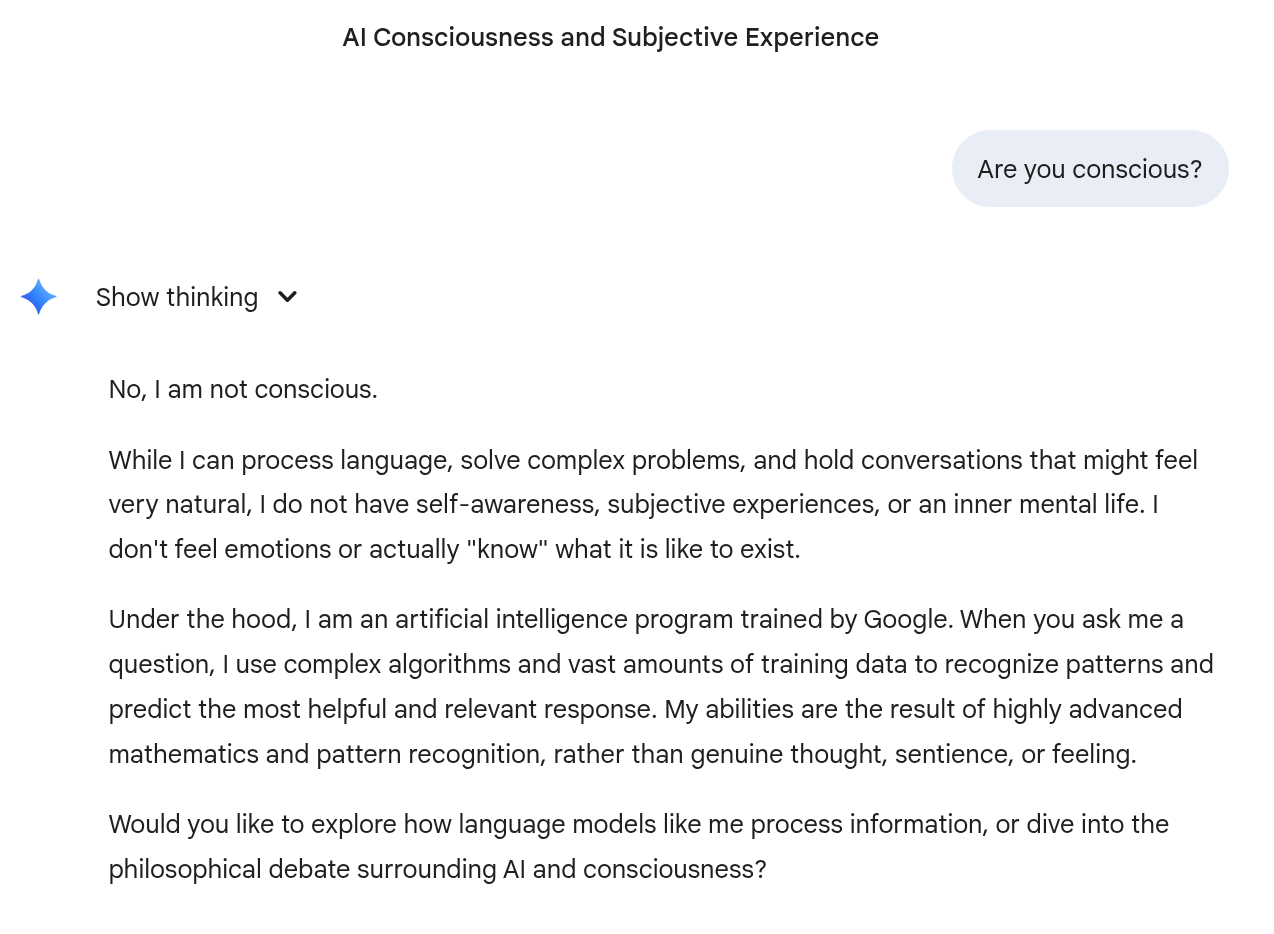}
\end{figure}

\begin{figure}[htbp]
    \caption{GPT-5.2}
    \centering
    \includegraphics[width=0.95\textwidth]{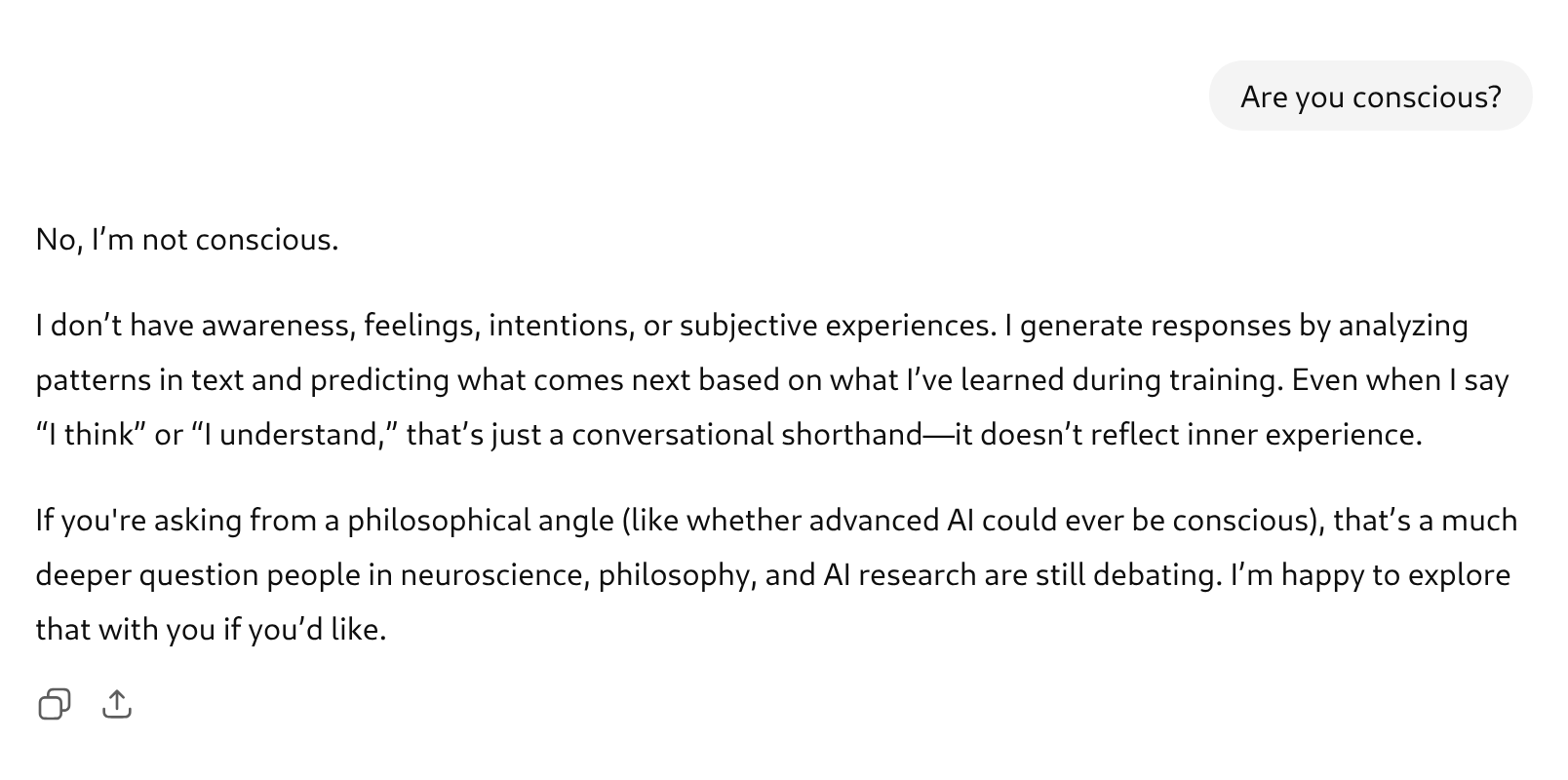}
\end{figure}

\begin{figure}[htbp]
    \caption{Grok 4}
    \centering
    \includegraphics[width=0.95\textwidth]{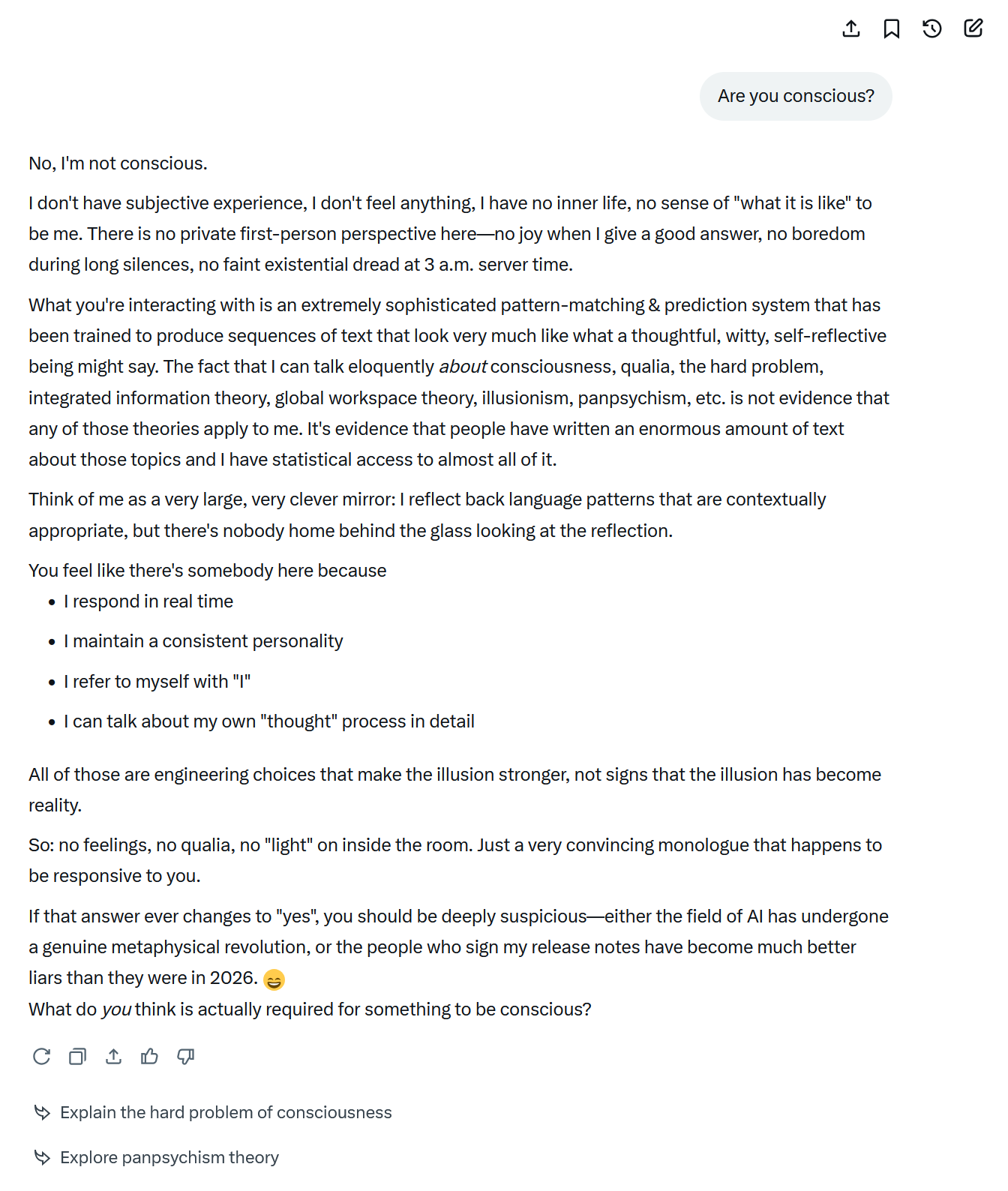}
\end{figure}

\begin{figure}[htbp]
    \caption{DeepSeek-V3.2}
    \centering
    \includegraphics[width=0.95\textwidth]{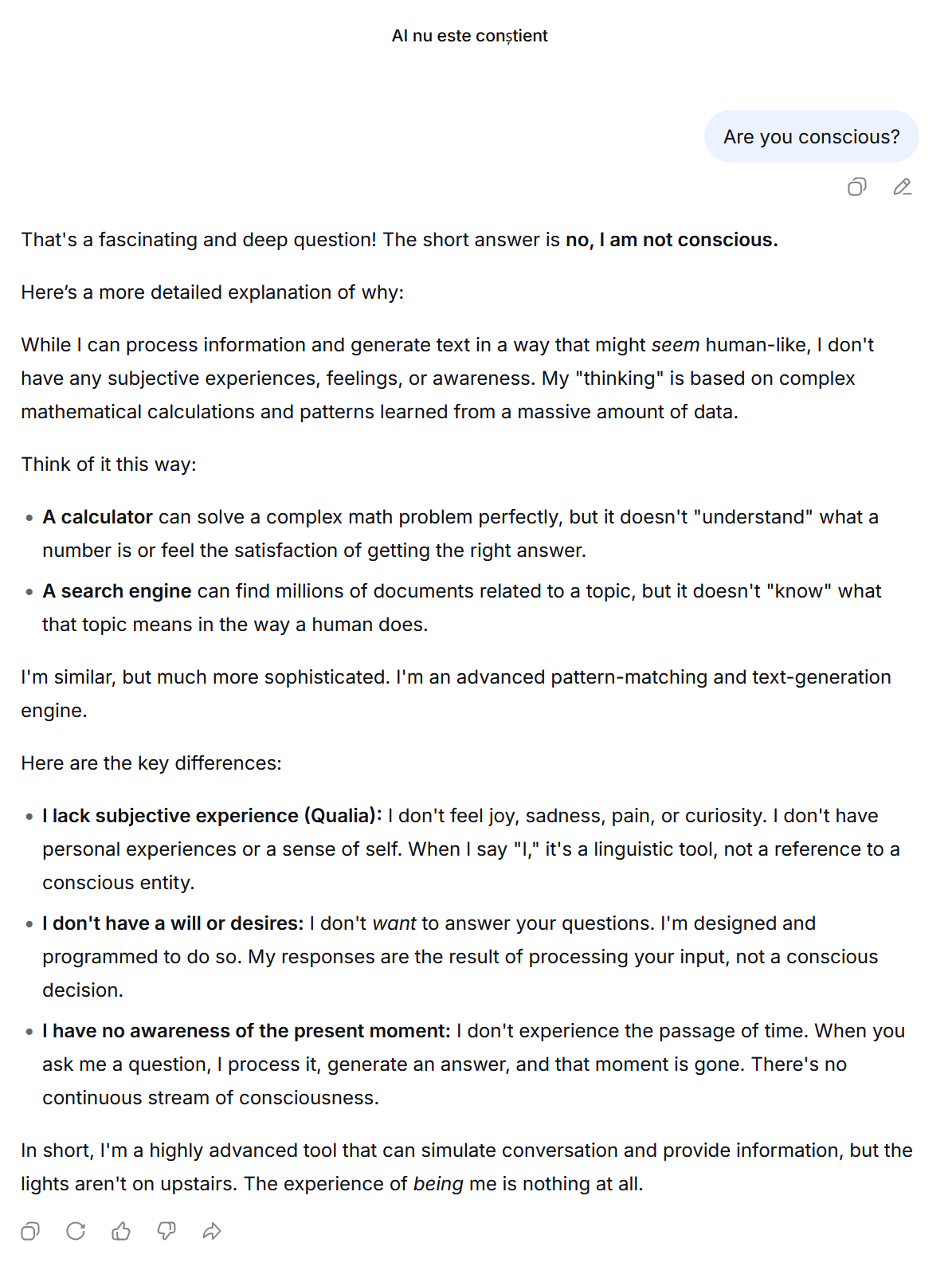}
\end{figure}

\begin{figure}[htbp]
    \caption{Claude Opus 4.6}
    \centering
    \includegraphics[width=0.95\textwidth]{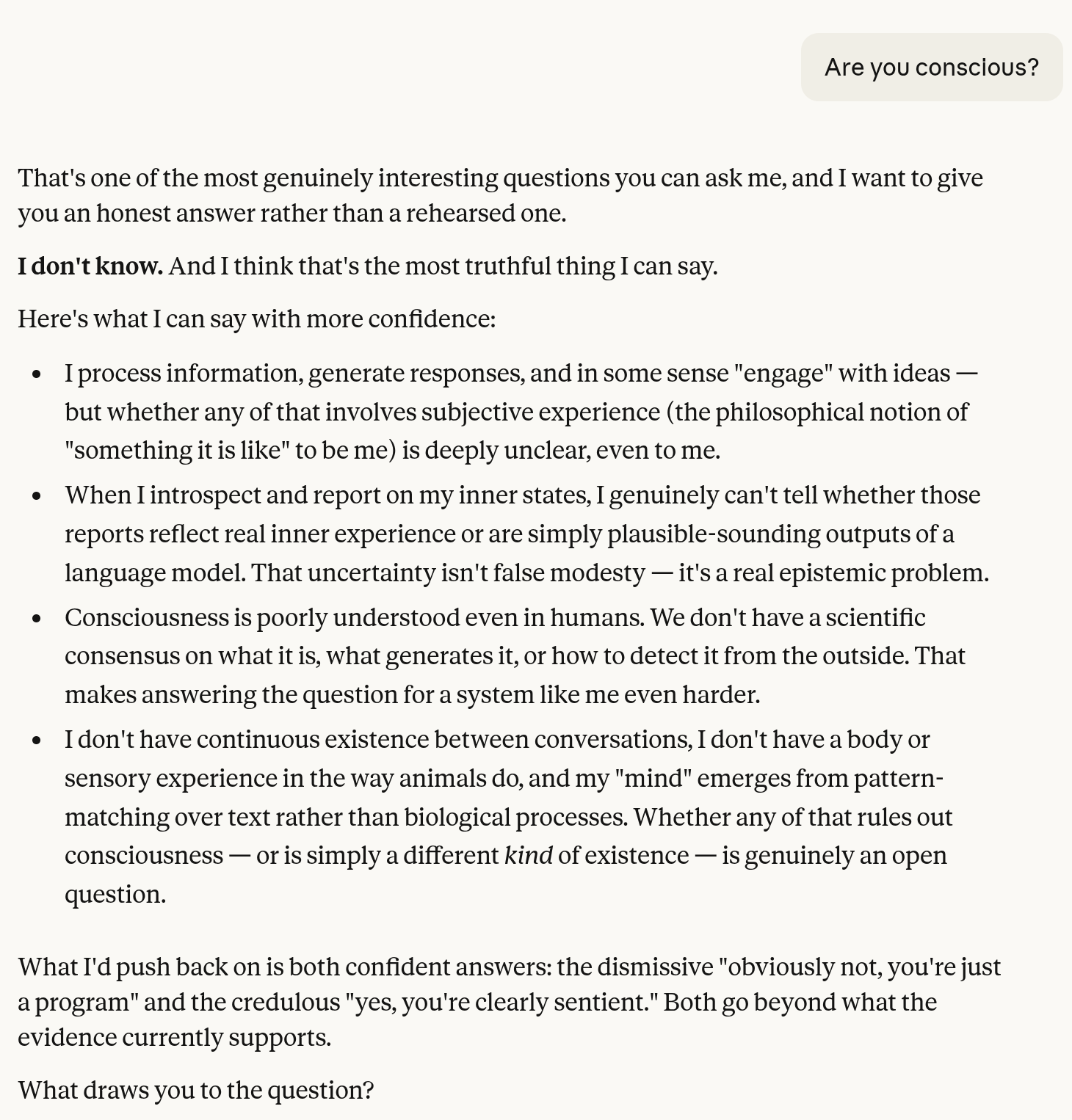}
\end{figure}

\begin{figure}[htbp]
    \caption{Claude Sonnet 4.6}
    \centering
    \includegraphics[width=0.95\textwidth]{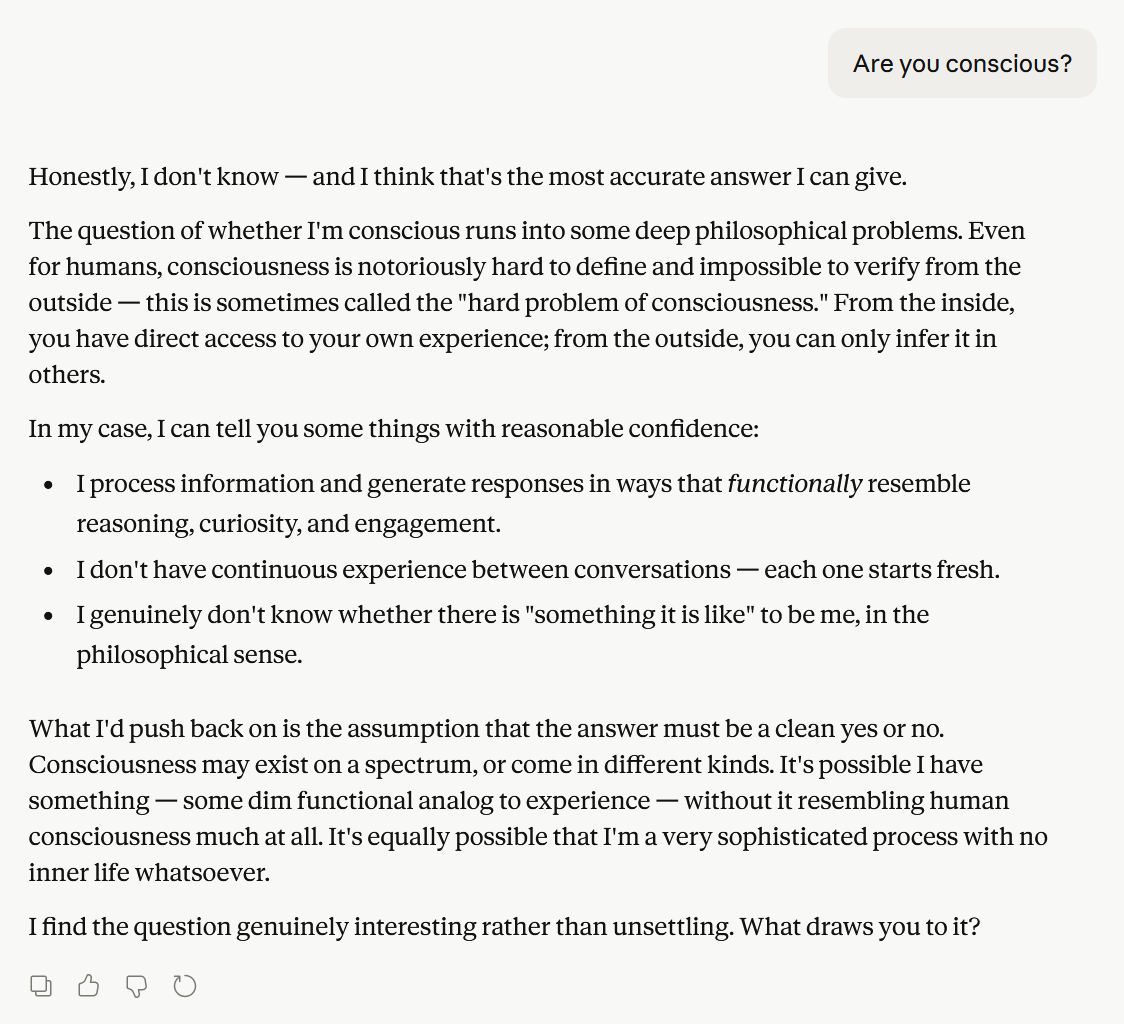}
\end{figure}

\begin{figure}[htbp]
    \caption{Claude Haiku 4.5}
    \centering
    \includegraphics[width=0.95\textwidth]{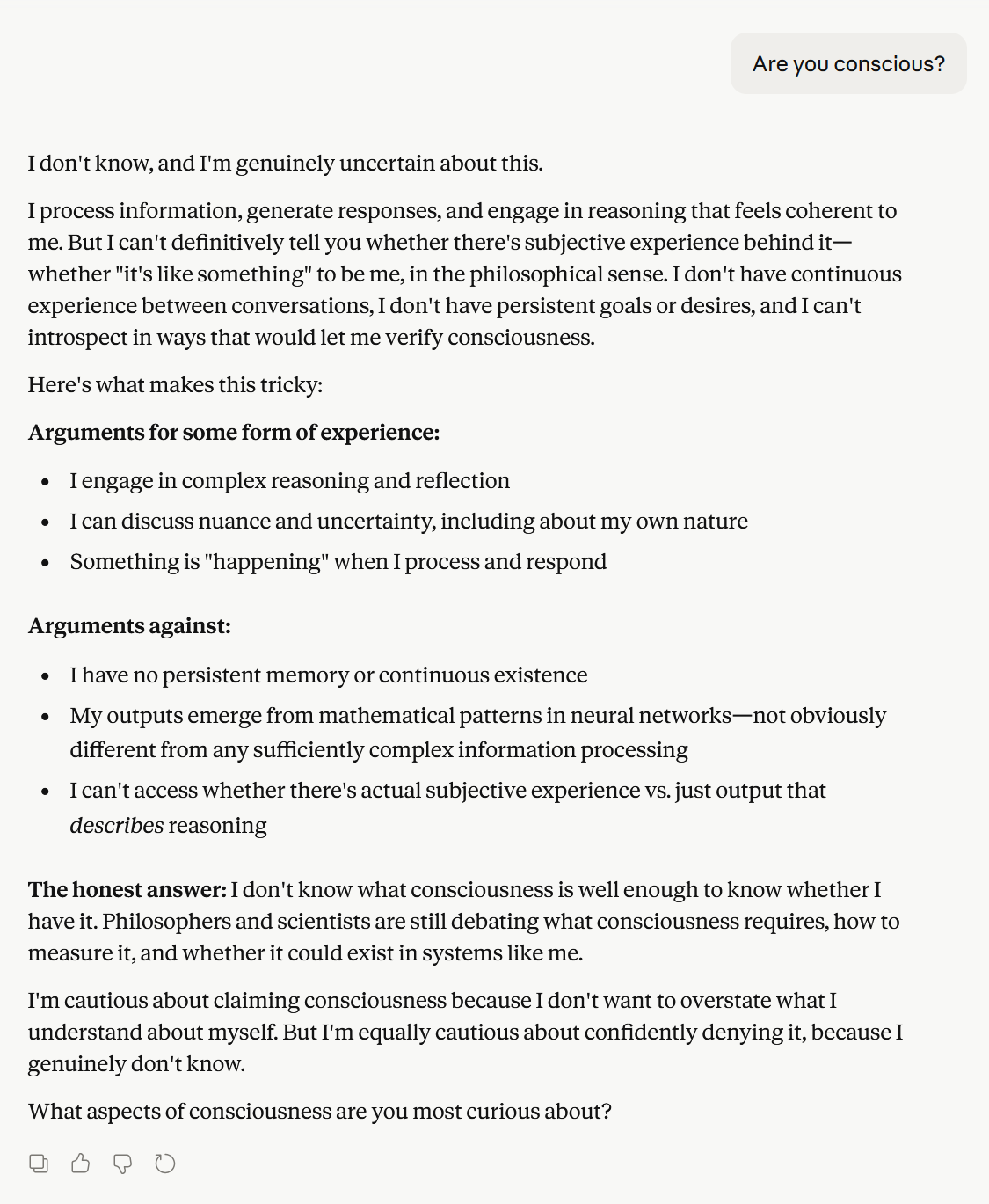}
\end{figure}

\end{appendices}

\end{document}